\documentclass[10pt]{JHEP3}

\usepackage{amsmath}
\usepackage{amsfonts}
\usepackage{graphicx}
\usepackage{bm}

\def\be{\begin{equation}}
\def\ee{\end{equation}}
\def\ba{\begin{eqnarray}}
\def\ea{\end{eqnarray}}
\def\bs{\begin{subequations}}
\def\es{\end{subequations}}
\def\cS{{\cal S}}
\def\cL{{\cal L}}
\def\B{\Box}
\def\l{\lambda}
\def\t{\theta}
\def\p{\partial}
\def\tphi{\tilde\phi}
\def\cO{{\cal O}}
\newcommand{\Eq}[1]{(\ref{#1})}

\title{Cosmological tachyon from cubic string field theory}
\author{Gianluca Calcagni\\
Astronomy Centre, University of
Sussex, Brighton BN1 9QH, United Kingdom\\
E-mail: \email{g.calcagni@sussex.ac.uk}}
\date{December 20, 2005}

\abstract{The classical dynamics of the tachyon scalar field of cubic string field theory is considered on a cosmological background. Starting from a nonlocal action with arbitrary tachyon potential, which encodes the bosonic and several supersymmetric cases, we study the equations of motion in the Hamilton--Jacobi formalism and with a generalized Friedmann equation, appliable in braneworld or modified gravity models. The cases of cubic (bosonic) and quartic (supersymmetric) tachyon potential in general relativity are automatically included. We comment the validity of the slow-roll approximation, the stability of the cosmological perturbations, and the relation between this tachyon and the Dirac--Born--Infeld one.}

\preprint{JHEP05(2006)012 \hspace{2cm} \hepth{0512259}}
\begin{document}


\section{Introduction}

Although string theory is still far from being completely understood, there has been compelling progress in the past decade, chiefly as regards its nonperturbative objects, the $Dp$-branes, and their role in determining the stable vacuum of the theory. In particular, the study of the open string tachyon mode of unstable $Dp$-branes has been pioneered by Sen \cite{sen5,sen6,sen7} via conformal field theory techniques. These and other methods (reviewed, e.g., in \cite{sen04}) all agree in the main results and show that as the tachyon rolls down towards the asymptotic minimum of the potential, the brane it lives on decays into a lower-dimensional brane or the closed string vacuum. In general, the tachyon dynamics can be described by a Dirac--Born--Infeld (in short, DBI) effective action  \cite{gar00,bers0,klu00,GHY}.

A particularly interesting approach is based on an attempt, called string field theory (SFT), to second-quantize the open bosonic string in a nonperturbative way.\footnote{Since the single string is a collection of modes which correspond to \emph{particle} fields, the theory of \emph{string} fields may be regarded as a `third quantization' scheme.} The starting point is the Chern--Simons-like action \cite{wi86a}
\be\label{SFT}
\cS=-\frac{1}{g_o^2}\int \left(\frac{1}{2\alpha'} \Phi* Q_B\Phi+\frac13\Phi*\Phi *\Phi\right),
\ee
where $g_o$ is the open string coupling constant (with dimension $[g_o^2]=E^{6-D}$), $\int$ is the path integral over matter and ghost fields, $Q_B$ is the BRST operator, * is the star product, and the string field $\Phi$ is a linear superposition of states in the Fock-space representation, whose coefficients correspond to the particle fields of the string spectrum. We refer to this as bosonic cubic SFT, or bosonic CSFT. (There is another string field theory, called boundary or background-independent SFT, which we will not consider here). The reader can find three of many excellent reviews on CSFT in \cite{sen04,ohm01,ABGKM}.

At the lowest truncation level, all particle fields in $\Phi$ are neglected except the tachyonic one, labeled $\phi(x)$ and depending on the center-of-mass coordinate $x$ of the string. That is to say, the Fock-space expansion of the string field is truncated so that, by virtue of the state-vertex operator isomorfism, $\Phi \cong |\Phi\rangle=\phi(x)|\!\downarrow\rangle$, where $|\!\downarrow\rangle$ is the ghost vacuum with ghost number $-1/2$.
Then at level $(0,0)$,\footnote{We recall that the \emph{level} of a state is the sum of the level
numbers of the creation operators acting on $|\!\downarrow\rangle$. At level $(L,M)$, the string field includes terms up to level $L$, while the action includes terms up to level $M$.} the action becomes \cite{KS1,KS2}, in $D=26$ dimensions and with metric signature $({-}{+}{\dots}{+})$, 
\be
\bar{\cS}_\phi=\frac{1}{g_o^2}\int d^D x \left[\frac{1}{2\alpha'}\phi(\alpha'\p_\mu\p^\mu+1)\phi-\frac{\l}{3}\left(\l^{\alpha'\p_\mu\p^\mu/3}\phi\right)^3-\Lambda\right],\label{tactmin}
\ee
where $\l=3^{9/2}/2^6\approx 2.19$, $\alpha'$ is the Regge slope, and Greek indices run from 0 to $D-1$ and are raised and lowered via the Minkowski metric $\eta_{\mu\nu}$. The tachyon field is a real scalar with dimension $[\phi]=E^2$. The extra constant $\Lambda$ does not contribute to the equation of motion for the scalar field but it does determine its dynamical behaviour. In particular, it corresponds to the $D$-brane tension which sets the height of the tachyon potential at the (closed-string vacuum) minimum to zero. This happens when $\Lambda=(6\l^2)^{-1}$, which is around $68\%$ of the brane tension; this value is lifted up when taking into account higher-level fields in the truncation scheme.

In order for CSFT to be a sensible candidate for a nonperturbative formulation of string theory, it has to reproduce the results of the conformal or boundary field theory approaches. In other words, the CSFT tachyon should be the same object as that in the latter cases (which for definiteness we shall identify with the DBI picture), sharing similar characteristics as regards rolling solutions, brane decay, and so on. With `same' object we mean that there should exist a field redefinition which maps tachyonic solutions obtained with different methods. Therefore it is important to study directly the CSFT action and its rolling tachyon solutions, which were inspected in \cite{MZ,yan02,FH1,FH2,FGN,CST} both analytically and numerically.

The DBI model has been considered also in cosmology \cite{gib02}, where the rolling homogeneous tachyon field can be regarded as an (or the) inflaton or dark energy field. The generalization to a cosmological context is natural because of the kind of evolution the DBI tachyon undergoes during its rolling down the potential: calling $\rho$ and $p$ its energy density and pressure, the field can behave as a dynamical cosmological constant, $-\rho\lesssim p < 0$. However, since its equation of state $p=w\rho$ is such that $-1\lesssim w<0$, the DBI scalar cannot decay faster than matter ($\rho\sim a^{-3}$, $w\sim 0$) after reheating, regardless the features of the latter. Therefore it is difficult to adopt the DBI tachyon as a model of dark energy. The problems worsen when imposing the tachyon to describe both the inflaton and quintessence field, because of the severe constraints from nucleosynthesis. The literature on the subject is rather extensive; for an overview see, for instance, \cite{CGT,PhD} and references therein.

The rolling CSFT solutions in flat spacetime have the key property to allow for positive values of the pressure after a pressureless phase similar to that of the DBI tachyon, which may open up the possibility to describe viable cosmological scenarios. Then, as for the DBI tachyon, it would be instructive to study the cubic tachyonic action on a curved background and, in particular, in a Friedmann--Robertson--Walker (FRW) spacetime. 

The goal of this paper is to present general arguments describing the behaviour of the CSFT cosmological tachyon. The action and covariant equations of motion on an arbitrary metric background are settled in sections \ref{setup} and \ref{eoms}, respectively, where the energy-momentum tensor for a CSFT tachyon with general potential is computed. In section \ref{cosmo} we consider several important topics which arise when the background metric is the FRW one. Section \ref{coseq} presents the cosmological equations, also in their Hamilton--Jacobi formulation, for either the general relativistic and braneworld case. The integration formula for the number of $e$-foldings is shown in section \ref{cosef}. Although we do not construct exact solutions of the equations of motion, some remarks on them are collected in section \ref{cosso}. The slow-roll appoximation is discussed in section \ref{cossr}, where it is shown that nonlocality requires a nontrivial refinement of the standard assumptions. Approximate solutions can be found in a quasi de Sitter limit, section \ref{cosds}, where however the action becomes purely local. Section \ref{cosst} is devoted to the issue of stability and the presence of ghost modes. We focus on the relation between the DBI and CSFT tachyons in section \ref{cosfr}. Discussion and prospects are in section \ref{disc}.

Before concluding the section, let us review the kind of potentials arising in CSFT. From now on we set $\alpha'=1$. 
When integrating out the other particle fields in the string field $\Phi$, at a given truncation level the effective tachyon potential acquires new terms \cite{ohm01}. In the bosonic CSFT, the tachyon potential $V$ at zero momentum has also quartic and higher-order contributions:
\be
V(\phi)=-\frac{\phi^2}{2}+\sum_{n\geq 4} a_n\phi^n.
\ee
In the general off-shell case, exponential differential operators will act on $\phi$. 

In superstring theory, tachyon modes appear on non-BPS branes (or $D$-$\bar{D}$ configurations, in which case the tachyon field is complex). So the dimension $D$ in the integral measure of the action is interpreted as the dimension $D=p+1$ of an unstable $Dp$-brane.\footnote{In the braneworld case, we assume that moduli fields from the extra directions, compactified or not, are stabilized. The volume of the compactified dimensions will be factored out from the action.} The supersymmetric version of open CSFT (henceforward CSSFT) is harder to formulate and there have been several attempts in this direction. The introduction of picture-changing operators allows for more or less natural generalizations of the noncommutative action, but also poses a number of problems not always easily solvable (we refer again to \cite{ohm01} for a review). Here we only quote the results for the tachyon effective potential in some CSSFT's. The first proposal, by Witten \cite{wi86b}, predicts a level 0 truncated tachyon potential which is quadratic and negative definite. At level $(1,2)$,\footnote{All the following expressions are at zero momentum and integration of the other particle fields is understood.} higher-order terms appear in the potential but do not improve its shape, $V(\phi)\sim -\phi^2/2-\phi^4/(1-16\phi^2)+\dots$ \cite{DR}. This potential has singularities and no minimum, and in general the theory suffers from tree-level contact divergences. 

For these and other reasons, other two candidates seem more promising. The first one is a modified (0-picture) CSSFT with a nonchiral, bilocal double-step operator \cite{PTY}, where contact divergences disappear. In this case the tachyon potential has two global minima \cite{AKBM}:
\be\label{bilV}
V(\phi)= -\frac{\phi^2}{4}+\frac{\l^{4/3}}{36}\phi^4,
\ee
at level (1/2,1). The coefficient of the quartic term has been computed also at level (2,6) but its magnitude ($\sigma=5053/69120\approx 0.073$) does not change significantly ($\l^{4/3}/36\approx 0.079$). A modified CSSFT with a chiral, local double-step operator \cite{AMZ1,AMZ2} gives the same $S$-matrices and different
off-shell effective potentials, but is less suitable for calculations.\footnote{The inequivalent dynamics in the Neveu--Schwarz sector of local and bilocal CSSFT's is perhaps an artifact of the level truncation approximation, which should not affect the full equations of motion in a `smooth' gauge \cite{are06}.} 

The second example is Berkovits' CSSFT \cite{ber95,ber99}. Again, the tachyon potential has a double-well shape \cite{ber00}:
\be\label{berV}
V(\phi)= -\frac{\phi^2}{4}+\frac{\phi^4}{2}+\dots,
\ee
where dots stand for higher level corrections which make the minima deeper. Since the qualitative features of this potential do not change when considering these terms, we shall not write them explicitly and refer to the original results \cite{BSZ,DR2,IN}.

Among other string field theories of interest we mention closed bosonic SFT \cite{zwi93} (see \cite{HMT} for an overview) and heterotic SFT \cite{OZ}. Presently, the above examples already give a good outlook of the CSFT zoology.

The cosmology of the CSFT tachyon is not a completely new subject and has been recently studied in \cite{are04,AJ} for an approximation of the modified CSSFT with bilocal double-step operator, where approximated solutions which are asymptotically de Sitter (i.e., with constant Hubble parameter) were found in four dimensions and for a standard Friedmann equation; below we shall discuss those results and point out some related caveats.


\section{Setup}\label{setup}

From the introduction, it is clear that there are several string models at stake, which predict different potentials according to the target symmetry and truncation level. In particular, the next-to-quadratic term in the tachyon effective potential is cubic in the bosonic string and quartic in the supersymmetric one. Within each string theory, one can include subdominant correction terms at higher truncation levels. These would be sufficient arguments for considering a model with a general potential. We would like also to add the following. To the best of our knowledge it is not obvious that, on a curved background, the tachyon potential should be one of those above, since the fundamental theory is formulated on a Minkowskian target spacetime. In some sense this objection applies to the whole concept of CSFT in a cosmological context as well, but the hope is, on one hand, to provide semi-quantitative insights to the problem of how string modes behave in an expanding universe and, on the other hand, to trigger further study of a relatively new aspect of such problem. This is also the point of view adopted in many works on the DBI cosmological tachyon, in which the tachyon potential is taken to be generic. In the concluding section we will do other comments on this point.

Gathering together all the pieces of information from the various string theories, we arrive at a compact formulation of the action to study. Coupling the tachyon to a metric $g_{\mu\nu}$, one has
\be\label{tact}
\cS_\phi=\int d^D x \sqrt{-g}\left[\frac12\,\phi(\B-m^2)\phi-U(\tphi)-\Lambda\right],
\ee
where $g$ is the determinant of the metric, $m^2$ is the squared mass of the scalar field (negative for the tachyon), and we have absorbed the open string coupling into $\phi$, so that the latter has dimension $[\phi]=E^{(D-2)/2}$ (later on, one can restore this factor by multiplying the pressure and energy density times $g_o^2$). The D'Alembertian operator is
\be\label{dal}
\Box \equiv \frac{1}{\sqrt{-g}}\,\p^\mu (\sqrt{-g}\,\p_\mu),
\ee
and we have defined
\be
\tphi\equiv \l^{\B/3}\phi.
\ee
The differential operator inside $\tphi$ can be expanded in an infinite series,
\be\label{lb}
\l^{\B/3}=\sum_{\ell=0}^{+\infty}\frac{(\ln\l)^\ell}{3^\ell \ell!} \B^\ell\equiv \sum_{\ell=0}^{+\infty}c_\ell \B^\ell,
\ee
where $\B^\ell$ is the $\B$ operator applied $\ell$ times. For future reference, note that $c_1=\ln\l/3=\ln 3^{3/2}-\ln 4\approx 0.26$. One can integrate this series (applied to $\phi$) term by term if both it is uniformly convergent and the individual terms are continuous. Continuity is guaranteed as long as $\phi(t)$ is a ${\cal C}^\infty$ function, while uniform convergence depends on how the D'Alembertian acts on $\phi(t)$. In the following we shall assume that both conditions are satisfied, keeping in mind that, once found an exact (or approximated) solution, one should perform a consistency check via a uniform convergence test as Abel's or Weierstrass'.

To the action (\ref{tact}) one must add the dynamical action for the graviton which, for sake of simplicity, we temporarily assume to be the Einstein--Hilbert one, so that the total action reads
\be
\cS=\cS_g+\cS_\phi,\qquad \cS_g=\frac{1}{2\kappa_D^2}\int d^D x \sqrt{-g}\, R,
\ee
where $\kappa_D$ is the effective gravitational coupling and $R$ is the target Ricci scalar. Although the term $\Lambda$ in the tachyonic action may be regarded as a cosmological constant, we shall include it in the definition for the tachyonic effective potential:
\be
\tilde V(\tphi) \equiv \tfrac{1}{2}m^2\phi^2+U(\tphi)+\Lambda.
\ee
This is not the potential of the scalar field since it contains differential operators; however, the potential energy of $\phi$ is included in it, $V(\phi) =\tilde V(\tphi)|_{\l=1}$. The zero-momentum quartic terms in the supersymmetric potentials above are modified into a potential $U$ for the `dressed' field $\tphi$. For instance, in \Eq{bilV} one has to replace
\be
\phi^4\to \left(\l^{\B/3}\tphi^2\right)^2.
\ee
The authors of \cite{AJK} approximated the potential to $\phi^4\to \tphi^4$, assuming that $\l^{\B/3}\tphi^2\approx \tphi^2$. For completeness and reasons which will become clear later, we shall not do so in the following.


\section{Covariant equations of motion}\label{eoms}

Let $U(\tphi)$ be the part of the tachyon effective potential which depends only on $\tphi$ and operators $\l^{p\B/3}$ acting on $\tphi^2$, where $p\in P \subseteq \mathbb{Z}\backslash\{0\}$ are integer numbers chosen in an appropriate set. With the symbol $U'$ we will indicate the variation of $U$ with respect to the tachyon:
\ba
U'&\equiv& \frac{\delta U}{\delta\phi}=\l^{\B/3}\tilde U',\\
\tilde U' &\equiv& \frac{\p U}{\p\tphi}+2\tphi\sum_{p\in P}\l^{p\B/3}D_pU,\\
D_pU &\equiv&\frac{\p U}{\p(\l^{p\B/3}\tphi^2)}.
\ea
Therefore, for a pure monomial potential with $P=\emptyset$,  
\ba
&&U(\tphi) = \frac{\sigma}{n}\,\tphi^n,\\
&& U'=\sigma \l^{\B/3}\tphi^{n-1},
\ea
where $\sigma$ is a constant. The bosonic CSFT is realized when
\be\label{boso}
m^2=-1,\qquad \sigma=\l,\qquad n=3,
\ee
while for $n=4$ one recovers the approximate model of \cite{AJK,are04,AJ}. For the modified CSSFT, $P=\{1\}$:
\ba
&&m^2=-1/2,\qquad U(\tphi) = \frac{\sigma}{4}\left(\l^{\B/3}\tphi^2\right)^2,\\
&&U'=\sigma\l^{\B/3}\left(\tphi\l^{2\B/3}\tphi^2\right),\qquad D_1 U =\frac{\sigma}{2}\l^{\B/3}\tphi^2.
\ea
The equation of motion for the tachyon is given by the field variation of the action, $\delta\cS_\phi/\delta \phi=0$:
\be\label{teom}
-(\B-m^2)\phi+U'=0\,.
\ee
Eventually one can recast eq. (\ref{teom}) in terms of $\tphi$,
\be\label{teom2}
-(\B-m^2)\l^{-(2/3)\B}\tphi+\tilde U'=0,
\ee
since an operator commutes with its functionals. In general, for a Lagrangian $\cL_\phi$ with an infinite number of D'Alembertian operators the equations of motion are
\be
\sum_{\ell=0}^{+\infty} \B^\ell\frac{\p \cL_\phi}{\p(\B^\ell\phi)}=0.
\ee
Also,
\be
\p_\mu\cL_\phi=\sum_{\ell=0}^{+\infty} (\p_\mu\B^\ell\phi)\frac{\p \cL_\phi}{\p(\B^\ell\phi)}.
\ee
Higher-derivative and nonlocal (i.e., containing infinitely many derivatives of the fields) Lagrangians in Minkowski spacetime were studied, for instance, in \cite{ost18,BFOW,EW2,NH,woo00} and \cite{LV,GKL,beri0,CHY,vol03,VV,GKR,vla05}.

The Einstein equations, obtained by varying the action with respect to the metric, read
\be
R_{\mu\nu}-\tfrac{1}{2}g_{\mu\nu}R=\kappa_D^2T_{\mu\nu},
\ee
where the energy momentum tensor is defined as
\be\label{set}
T_{\mu\nu}\equiv-\frac{2}{\sqrt{-g}}\frac{\delta \cS_\phi}{\delta g^{\mu\nu}}.
\ee
Note that this definition is used even in Minkowski spacetime due to the difficulty in defining the pressure unambiguously. In such a case one promotes the Minkowski metric to a general one and varies with respect to it, setting then $g_{\mu\nu}=\eta_{\mu\nu}$ in the energy tensor. However, here we adopt a different perspective and the promotion to a general background acquires physical meaning. Then eq.~\Eq{set} is not just a trick but the covariant definition of the energy-momentum tensor in a general relativistic theory.

In order to deal with the variation $\delta\tphi/\delta g^{\mu\nu}$, one can follow two different approaches. The first one is to consider the expansion of $\tphi$, eq. \Eq{lb}. For each element of the series and two scalars $f_1$ and $f_2$, one can use the relation, valid inside the integral after an integration by parts,
\be
f_1(\delta \B^\ell)f_2=\sum_{j=0}^{\ell-1} (\B^j f_1)(\delta\B) (\B^{\ell-1-j}f_2);
\ee
to this, one applies the relation
\be\label{dB}
q_1\frac{\delta\B}{\delta g^{\mu\nu}} q_2 =\frac12 g_{\mu\nu} (q_1\B q_2+\p_\alpha q_1 \p^\alpha q_2)-(\p_\mu q_1)( \p_\nu q_2),
\ee
where $q_1=\B^j f_1$, $q_2=\B^{\ell-1-j}f_2$, and we have used eq.~(\ref{dal}) together with $\delta\sqrt{-g}=-\tfrac12\sqrt{-g}\,g_{\mu\nu}\delta g^{\mu\nu}$. Then the energy-momentum tensor is
\be
T_{\mu\nu} =\p_\mu\phi\p_\nu\phi-g_{\mu\nu}\left(\tfrac12\p_\alpha\phi\p^\alpha\phi+\tilde V\right)+\tilde T_{\mu\nu},
\ee
where
\bs\label{Tmunu}\ba
\tilde T_{\mu\nu} &=& \sum_{\ell=1}^{+\infty}c_\ell\sum_{j=0}^{\ell-1}\tilde T_{\mu\nu}^{(\ell,j)},\\
\tilde T_{\mu\nu}^{(\ell,j)} &\equiv& g_{\mu\nu}\left[(\B^j\tilde U')(\B^{\ell-j}\phi)+(\p_\alpha\B^j\tilde U')(\p^\alpha\B^{\ell-1-j}\phi)\right]-2(\p_\mu\B^j\tilde U')(\p_\nu\B^{\ell-1-j}\phi)\nonumber\\
&&+\sum_{p\in P}p^\ell\left[g_{\mu\nu}(\B^j D_pU)(\B^{\ell-j}\tphi^2)+g_{\mu\nu}(\p_\alpha\B^jD_pU)(\p^\alpha\B^{\ell-1-j}\tphi^2)\right.\nonumber\\
&&\qquad\left.-2(\p_\mu\B^jD_pU)(\p_\nu\B^{\ell-1-j}\tphi^2)\right].\label{equa}
\ea\es
Equation (\ref{Tmunu}), which is written in a fully covariant way, agrees with the results of \cite{yan02} only in the particular case of Minkowski spacetime and cubic potential. When $P=\emptyset$, only the first line of eq. \Eq{equa} survives. Although it is not apparent, because of the symmetry of the Einstein tensor also $\tilde T_{\mu\nu}$ is symmetric in the indices.

Another method is to consider the relation, valid inside the $D$-dimensional integral and modulo total derivatives,
\be
f_1(\delta e^{c_1 \B})f_2=c_1\int_0^1 ds (e^{c_1 s\B}f_1)(\delta\B)[e^{c_1(1-s) \B}f_2],
\ee
and then use again eq.~\Eq{dB} with $q_1=\l^{s\B/3}f_1$ and $q_2=\l^{(1-s) \B/3}f_2$, giving
\ba
\tilde T_{\mu\nu} &=& c_1\int_0^1 ds\, \tilde T_{\mu\nu}^{(s)},\\
\tilde T_{\mu\nu}^{(s)}&\equiv& g_{\mu\nu}\left[(\l^{s\B/3}\tilde U')(\B\l^{-s\B/3}\tphi)+(\p_\alpha \l^{s\B/3}\tilde U')(\p^\alpha \l^{-s\B/3}\tphi)\right]-2(\p_\mu\l^{s\B/3}\tilde U')(\p_\nu \l^{-s\B/3}\tphi)\nonumber\\
&&+\sum_{p\in P}p
\left\{g_{\mu\nu}(\l^{ps\B/3}D_pU)[\B\l^{p(1-s)\B/3}\tphi^2]\right.+g_{\mu\nu}(\p_\alpha \l^{ps\B/3}D_pU)[\p^\alpha \l^{p(1-s)\B/3}\tphi^2]\nonumber\\
&&\qquad \left.-2(\p_\mu\l^{ps\B/3}D_pU)[\p_\nu \l^{p(1-s)\B/3}\tphi^2]\right\}.
\ea
When $P=\emptyset$, only the first line of the last equation survives. Every differential operator acts on all the right hand side within each parenthesis.


\section{Cosmology}\label{cosmo}

As a particular metric we choose the flat FRW one, which encodes the cosmological principle. In synchronous gauge, it reads
\be
ds^2=-dt^2+a^2(t)\,dx_i dx^i,
\ee
where $a(t)$ is the scale factor characterizing the physical size of the universe and Latin indices run over spatial coordinates. The world-volume of a FRW $Dp$-brane varies in time as $a^p(t)$. On this background, the energy of the tachyon will not be constant any longer, nor its evolution will be clearly interpreted as a condensation/brane decay mechanism.

The D'Alembertian operator on an homogeneous field becomes $\B=-\p_t^2-(D-1)H\p_t$, where $H\equiv \dot{a}/a=\p_t a/a$ is the Hubble parameter; its inverse marks the size of the causal horizon. We assume to be in an expanding universe, $H>0$. 


\subsection{Friedmann and Hamilton--Jacobi equations}\label{coseq}

The energy-momentum tensor yields the energy density and pressure of the tachyon field $\phi$:
\ba
\rho &=& -T_0{}^0= \frac{\dot{\phi}^2}{2}(1-\cO_2)+\tilde V -\cO_1,\label{rho}\\
p &=& T_i{}^i=\frac{\dot{\phi}^2}{2}(1-\cO_2)-\tilde V+\cO_1\label{pres}
\ea
(no trace over the index $i$). The functions $\cO_i$ are, in the series and integral representations,
\bs\label{Os}\ba
\cO_1 &=& \sum_{\ell=1}^{+\infty}c_\ell\sum_{j=0}^{\ell-1}
\Big[(\B^j\tilde U')(\B^{\ell-j}\phi)\vphantom{\sum_P}+\sum_{p\in P}p^\ell(\B^j D_pU)(\B^{\ell-j}\tphi^2)\Big]\\
&=& c_1\!\int_0^1 ds\, \Big\{(\l^{s\B/3}\tilde U')(\B\l^{-s\B/3}\tphi)+\sum_{p\in P}p\, (\l^{ps\B/3}D_pU)[\B\l^{p(1-s)\B/3}\tphi^2]\Big\},\\
\cO_2 &=& \frac{2}{\dot{\phi}^2} \sum_{\ell=1}^{+\infty}c_\ell\sum_{j=0}^{\ell-1}\Big[(\B^j\tilde U')^.(\B^{\ell-1-j}\phi)^.+\sum_{p\in P}p^\ell(\B^j D_pU)^.(\B^{\ell-1-j}\tphi^2)^.\Big]\\
&=& \frac{2c_1}{\dot{\phi}^2}\int_0^1 ds\, \Big\{(\l^{s\B/3}\tilde U')^.(\l^{-s\B/3}\tphi)^.+\sum_{p\in P}p\, (\l^{ps\B/3}D_pU)^.[\l^{p(1-s)\B/3}\tphi^2]^.\Big\}.
\ea\es
Thus the tachyon behaves as a perfect fluid with equation of state $p=w\rho$, $w$ being a function of time, and continuity equation $\nabla^\nu T_{\mu\nu}=0$, that is,
\be\label{cont}
\dot{\rho}+(D-1)H(\rho+p)=0.
\ee 
Alternatively, one might split the fluid into two components, $\rho=\rho_{\rm KG}+\tilde\rho$, describing a Klein--Gordon field sourced by some term enclosing higher derivatives. However, these components are not independent, nor presumably they would help in gaining further insight in the problem. 

Equations \Eq{Os} are not very transparent and it is helpful to consider the example of the bosonic tachyon:
\ba
\cO_1 &=& \l \sum_{\ell=1}^{+\infty}c_\ell\sum_{j=0}^{\ell-1}(\B^j\tphi^2)(\B^{\ell-j}\phi)\qquad\qquad\\
&=& \l c_1\!\int_0^1 ds\, (\l^{s\B/3}\tphi^2)(\B\l^{-s\B/3}\tphi)\,,\\
\cO_2 &=& \frac{2\l}{\dot{\phi}^2} \sum_{\ell=1}^{+\infty}c_\ell\sum_{j=0}^{\ell-1}(\B^j\tphi^2)^.(\B^{\ell-1-j}\phi)^.\\
&=&  \frac{2c_1\l}{\dot{\phi}^2} \int_0^1 ds\, (\l^{s\B/3}\tphi^2)^.(\l^{-s\B/3}\tphi)^.\,.
\ea
An apparent feature of the CSFT tachyon is that neither its energy density nor pressure are positive definite (the bosonic potential is even unbounded from below). Also, the pressure eq.~(\ref{pres}) can be greater that 0, in contrast with the DBI tachyon. This might overcome the problems of fine tuning typical of the cosmological DBI tachyon explained in the introduction. 

The Friedmann equation in four conformally flat dimensions reads
\be\label{FRW}
H^2=\frac{\kappa_4^2}{3}\,\rho,
\ee
and provide, together with the continuity equation, all the necessary information to find, at least in principle, a solution $\{a(t),\phi(t)\}$, which must satisfy the weak energy condition $\rho\geq 0$.\footnote{We should say that the Friedmann equation \Eq{FRW} is deceptively simple. The energy density contains $H$ and all its time derivatives through the functions $\cO_i$.} It is easy to extend the analysis to a generalized cosmological equation
\be\label{FRW2}
H^{2-\theta}=\beta_\t^2\rho,
\ee
where $\theta$ is a constant and $\beta_\t$ encodes the gravitational coupling. This kind of evolution arises asymptotically in high-energy braneworld scenarios and modified gravity theories. A detailed discussion on this approach and its derivation from an effective gravitational action can be found in \cite{PhD,hol}. The Einstein--Hilbert case is recovered for $\theta=0$.

In the Hamilton--Jacobi formalism the field $\phi$ plays the role of cosmic time provided $\dot{\phi}$ does not change sign, a standard situation during inflation.\footnote{Even if the solution $\phi(t)$ is not monotonic on its domain, one can limit the discussion to any finite time interval where this is the case. Then the found solutions can be joined together.} Differentiating eq.~\Eq{FRW2} with respect to $t$ and combining eqs. \Eq{rho}, \Eq{pres}, and \Eq{cont} one gets
\be\label{hj1}
\frac{H'}{H^\t} =-\frac{3q}{2}\,\beta_\t^2\dot{\phi}\,(1-\cO_2),
\ee
where $H=H(\phi)$, $q\equiv 2(2-\theta)^{-1}$, and primes denote differentiation with respect to $\phi$. The first slow-roll (SR) parameter $\epsilon\equiv -\dot{H}/H^2$ can be written, assuming $\cO_2\neq 1$ at any time, as
\ba
\epsilon &=&-\frac{H'}{H^2}\dot{\phi}=\frac{2}{3q\beta_\t^2}\frac{H'^2}{H^{2+\t}}\,(1-\cO_2)^{-1},\label{epsi}\\
&=&\frac{3q}{2}\beta_\t^2\,{H^{\t-2}}\dot{\phi}^2\,(1-\cO_2),\label{ep}
\ea
which yields, via eqs.~\Eq{FRW2} and \Eq{rho},
\be\label{hj2}
\left(1-\frac{\epsilon}{3q}\right)\frac{H^{2-\t}}{\beta_\t^2}+\cO_1-\tilde V=0,
\ee
or
\be\label{hj3}
\frac{2}{(3q\beta_\t^2)^2}\left(\frac{H'}{H^\t}\right)^2-(1-\cO_2)\left(\frac{H^{2-\t}}{\beta_\t^2}+\cO_1-\tilde V\right)=0.
\ee
The expressions \Eq{hj1} and \Eq{hj3} are the Hamilton--Jacobi equations. Sometimes the latter is recast as an equation in terms of $H(\phi)$ and $a(\phi)$, using $\epsilon=-aH'/(a'H)$:
\be\label{hjy1}
a'H'=-\frac{3q}{2}\beta_\t^2\,aH^{1+\t}(1-\cO_2).
\ee
Equations \Eq{teom} and \Eq{cont} provide a relation between $\cO_1$ and $\cO_2$:
\be
\cO_1'+\frac{\dot{\phi}^2}{2}\,\cO_2'=\tilde V'\cO_2,
\ee
where $\cO_i(\phi)$, $\tphi(\phi)$, and $\tilde V(\phi)$ are thought as functions of the new time coordinate. 

Equation \Eq{epsi} implies that an expanding universe superaccelerates if, and only if, $\cO_2$ is greater than 1:
\be
\epsilon<0\qquad \Leftrightarrow\qquad \cO_2>1,
\ee
provided $q$ is positive as in general relativity ($q=1$) and Randall--Sundrum ($\t=1$, $q=2$) or Gauss--Bonnet ($\t=-1$, $q=2/3$) braneworld scenarios. In this case, looking at eqs.~\Eq{rho} and \Eq{pres}, the tachyon behaves also as a ghost field (or phantom, in cosmological language), with a negative-definite effective kinetic term. This property may have consequences for dark energy phenomenology, as was already noted in \cite{are04,AJ}. Remarkably, this condition does not depend upon the type of potential and Friedmann equation governing the cosmological evolution, although the time dependence of $\phi$ and $H$ will. 

In general, one can identify three behaviours in the equation of state of the tachyon: \emph{Pseudo slow roll} ($w\sim -1$) when (Ia) $\dot\phi^2\ll \tilde V$ and $\dot\phi^2|\cO_2|\ll |\cO_1|$ or (Ib) $|\dot{\phi}^2(1-\cO_2)| \ll |\tilde V-\cO_1|$ or (Ic) $\cO_2\sim 1$; \emph{Pseudo kinetic regime} ($w\sim 1$, beyond the DBI barrier) when (IIa) $\dot\phi^2\gg \tilde V$ and $\dot\phi^2|\cO_2|\gg |\cO_1|$ or (IIb) $|\dot{\phi}^2(1-\cO_2)| \gg |\tilde V-\cO_1|$ or (IIc) $\tilde V\sim \cO_1$; (III) \emph{Pseudo phantom regime} ($w\lesssim -1$) when $\cO_2>1$.

When $\l=1$ in all the above equations, $\cO_i=0$ and one recovers the results for an ordinary scalar field \cite{PhD}.


\subsection{${\bm e}$-foldings}\label{cosef}

The number of $e$-foldings from the time $t$ to $t_*$ is $N(t)\equiv \int^{t_*}_t H(t') dt'$, which can be written as
\ba \label{N}
N(t) &=&-\frac{3q}{2}\beta_\t^2\int_{\phi(t)}^{\phi_*}d\phi\, (1-\cO_2)\frac{H^{\t+1}}{H'}\\
&=& \pm\int_{\phi(t)}^{\phi_*}d\phi\left[\frac{3q}{2}\beta_\t^2\, (1-\cO_2) \frac{H^{\t}}{\epsilon}\right]^{1/2},
\ea
where the sign ambiguity in the last equation is related to the sign of $\dot{\phi}$. If inflation occurs, $t_*$ is interpreted as the time when it ends.


\subsection{Cosmological solutions?}\label{cosso}

In Minkowski spacetime, an incomplete analytic solution for the bosonic equation of motion was calculated via the following procedure \cite{FGN}: ($i$) Treat $\ln\l$ as a free evolution parameter, so that the domain of the tachyon field $\phi(\ln\l,t)$ is contained in $\mathbb{R}^2$, ($ii$) Find an exact solution for $\l=1$ and interpret it as the initial condition for an evolution equation with respect to $\l$, ($iii$) Evolve the equation of motion to find its solution, valid for $\l<1$, and ($iv$) Analytically continue this solution to the region $\l>1$. The solution thus found is not exact but approximate, although with very high accuracy. Its energy density is constant, $\rho=\Lambda$, since the brane volume is. The field starts from the maximum of the potential at $t\to-\infty$, rolls down to the minimum (which is not at infinity as in the DBI case) and past it up to $t=0$, where however the solution becomes singular \cite{gri}. The actual evolution of the pressure depends on the value of $\Lambda$, but it can be set so that it goes from $-1$ at the maximum to 0 at the minimum, becoming positive between this and the breaking point. At this point there is solid numerical evidence that the dynamics experiences wild oscillations. This can be seen by expanding the field as an exponential series, $\phi(t)=\sum_{n} b_n e^{nt}$. The solution turns out to be accurate, at least up to some time after the beginning of the oscillatory regime, already when the series is truncated at $n=6$ \cite{CST}. The analytic behaviour of \cite{FGN} is in agreement with the numerical result \cite{CST}.

In the cosmological case the above scheme seems pretty hard to be developed. Actually, there is no analytic solution even in the case $\l=1$ for any $\t$ of interest and both the quadratic+cubic and quadratic+quartic potentials (compare Table 2.3 in \cite{PhD}), although a numerical solution is still achievable.

The diffusion equation found in \cite{FGN}, $\p\phi(\ln\l,t)/\p\ln\l=-\p_t^2\phi(\ln\l,t)$ is quite similar to that of the $1+1$ Hamiltonian formalism developed in \cite{LV,GKL,beri0,CHY,GKR}, which allows for a well-defined quantization. This formulation might be the most convenient to study the cosmological solutions, although it is not yet clear to us how to generalize the procedure to the covariant case. 

Instead of $\ln\l$ one might consider the coupling constant $\sigma$ of $U(\tphi)$ as evolving parameter, which would be a natural choice since it does change when increasing the expansion level. One can also start from the observation that $\sigma$ is typically smaller than 1, so that a perturbative expansion in $\sigma$ can be applied \cite{EW2,LV,CHY}.

It will be crucial to understand whether the Hubble friction term is able to damp the wild oscillations of the bosonic case; this would let the tachyon die down the minimum, eventually reheating the universe in a conventional way (that is, with no need of gravitational particle production or curvaton rescue). The bizarre oscillations sketched above are due to the infinite series of derivatives acting on the field, and only special kinds of friction terms might compensate the exponential kinetic factors. We do not solve this issue here. Rather, we ask which requirements the derivatives of the Hubble parameter should satisfy in an inflationary slow-roll regime.


\subsection{Localized action and extended SR approximation}\label{cossr}

The cosmological equations of motion are nonlinear and involve all the derivatives $\phi^{(n)}$ and $H^{(n)}$ of the field and Hubble parameter, respectively. There is an infinite number of initial conditions $\{\phi^{(n)}(t_0),H^{(n)}(t_0)\}$ to be specified in order to find a solution. As noted in \cite{MZ}, this is very different from the Cauchy problem for a finite number of initial conditions, since to find all of them would correspond to know the Taylor expansion of the solution $\{\phi(t),H(t)\}$ itself.\footnote{Another way to state this is that, in most cases where the Lagrangian is nonlocal, there is no theorem of existence and uniqueness of solutions \cite{LV}.} Therefore, rather than choosing arbitrary conditions and evolve from them, for a given potential one has to reconstruct the only consistent set of Taylor coefficients.

Moreover, when perturbing a solution $H_0(\phi)\to H(\phi)=H_0(\phi)+\delta H(\phi)$ and linearizing the Hamilton--Jacobi equation \Eq{hj3}, one gets an equation for $\delta H$ with infinite time derivatives. Then it seems quite hard to say anything about the inflationary attractor either. Roughly speaking, nonlocal theories are rather unfriendly to the inflationary paradigm, that is, while the latter tends to erase any memory of the initial conditions, the formers do preserve this memory in a way or another. In noncommutative cosmology (see \cite{PhD} for details and references) the string scale breaks the scale invariance of the power spectrum, by selecting a particular wavelenght, and the insensitivity of the gauge invariant perturbations from initial conditions is in part lost. In the model we are considering, the presence of all the Taylor coefficients of the fields in the equations of motion alters the Cauchy problem in the way described above, so that there is much less room for arbitrariness in the choice of a suitable solution.

For numerical purposes, one is forced to truncate the exponential series of D'Alembertians at some, possibly large, order. In a similar way, the infinite SR tower is typically truncated at some level and only the lowest order terms are kept, while the smaller higher-order terms are neglected. At first glance, a truncation of the differential operators might help considerably. In fact, when the series of D'Alembertians is finite, there is also a finite number of SR parameters in the equations of motion. This is a situation analogous to that of a theory with finite derivatives as in the cases of a normal or DBI scalar field in $f(R, R_{\mu\nu}R^{\mu\nu},R_{\mu\nu\sigma\tau}R^{\mu\nu\sigma\tau})$ gravities. Then one is allowed to keep only the leading SR terms.\footnote{To some degree, the SR structure of the equations of motion depends on the definition of the parameters themselves. However, the number of independent SR parameters is always finite in the ordinary equations of motion. For instance, the SR parameter $\eta$ for the DBI tachyon can be defined in a way such that it actually contains a factor $(1-\epsilon)^{-1}$, which can be expanded as $1+\epsilon+\epsilon^2+\dots\,$. To keep only the leading term would truncate an infinite series, but of only one SR parameter. See eq. (2.57) in \cite{PhD} and related discussion.} 

We must note that the series and slow-roll truncations are \emph{not} equivalent, and while the series truncation always permits the standard SR approximation, the converse is not true. Before truncating the series, the infinite set of SR parameters appears in the equations, since $\B^n\phi\sim O[\epsilon^{n-(1/2)}]$, and one might not be authorized to discard most of these terms, whose infinite sum may be comparable to the other surviving pieces. The symbol $O(\epsilon^n)$ collects all the terms of the SR tower from the $n$th order: $\epsilon^m$, $\eta^m$, $\epsilon^k\eta^{m-k}$\dots, for $m\geq n$; see below for the definitions. The above power counting, valid for a quasi de Sitter regime $H\approx{\rm const}$, follows from the fact that each $\B$ raises the time derivatives by one (from the dominating friction term), and each time derivative raises the SR order by one, since $\dot{\epsilon}\sim  O(\epsilon^2)$. But  $\B\phi=H\dot{\phi}(\eta-3)\sim O(\epsilon^{1/2})$ from eq.~\Eq{ep}, hence the result.

Therefore, the SR approximation is a priori feasible only after truncating the derivative series at a finite order
\be\label{appr}
\l^{\B/3}\to \sum_{\ell=0}^{\ell_{\rm max}}c_\ell \B^\ell,
\ee
no matter how large $\ell_{\rm max}$ is (in \cite{MZ}, $\ell_{\rm max}\sim O(50)$). The truncation procedure, eq.~\Eq{appr}, might lead to solutions rather different from those of the original theory, since the truncation loses an infinite number of constraints and hence allows for an infinite number of spurious solutions. Anyway, there is evidence of good convergence of the truncated theory in the Minkowski case \cite{MZ,CST}. However, we stress that there is a fundamental reason why the localized (i.e., truncated) theory should be regarded as an object different from the nonlocal one. Namely, in the first case the scalar propagator has multiple poles and at least one of them has negative residue, while in the second case the theory can be ghost-free \cite{BMS}.

We can take another perspective and ask what happens if the SR approximation is, somehow, guaranteed a priori. It is useful to introduce the second and third SR parameters
\be
\eta\equiv-\frac{\ddot{\phi}}{H\dot{\phi}},\qquad \xi^2\equiv \frac{\dddot{\phi}}{H^2\dot{\phi}}-\eta^2.
\ee
In a not too strong SR regime, both $\epsilon$ and $\eta$ are very small and almost constant, so that the scale factor evolves approximately as a power law, $a\sim t^{1/\epsilon}$. Then one can show that, for exactly constant $\epsilon$ and $\eta$,
\be
\tphi =-\sum_{\ell=0}^{+\infty}c_\ell[3-2(\ell-1)\epsilon-\eta](\epsilon t)^{1-2\ell}\dot{\phi}\prod_{n=0}^{\ell-2}(3-2n\epsilon-\eta)[(2n+1)\epsilon+\eta].
\ee
Since $\eta\sim\epsilon\ll 1$ implies $\dot{\phi}\sim t$, $\B^\ell\phi\sim f(\ell)t^{2-2\ell}$, where $f(\ell)$ does not grow faster that $t^{2\ell-2}$ with respect to $\ell$. Therefore it would seem that, if the SR approximation is valid and sufficiently effective, the exponential operators give a finite result aymptotically in time. However, the magnitude of $\eta$ turns out to be of difficult assessment once the Hubble parameter is known, since eq.~\Eq{ep} contains infinitely many factors of $\dot{\phi}$, and the above rough estimate actually relies on a condition much stronger that the traditional SR one. In fact, the evolution equations for $\epsilon$ and $\eta$, using eq.~\Eq{ep}, are
\ba
\dot{\epsilon}&=& H\epsilon\,[(2-\t)\epsilon-2\eta-\gamma],\label{ee1}\\
\dot{\eta} &=& H(\epsilon\eta-\xi^2),\label{ee2}\\
\gamma &\equiv& \frac{\dot{\cO}_2}{H(1-\cO_2)}.\label{ee3}
\ea
While eq.~\Eq{ee2} is the same as for a normal scalar field, eq.~\Eq{ee1} depends on both the type of gravity one is living in (through the parameter $\t$) and a new parameter related to $\cO_2$. Then, in order for $\epsilon$ and $\eta$ to be constant, one has to impose simultaneously
\ba
&&\epsilon\ll 1,\label{condsr1}\\
&&\eta\ll 1,\qquad \xi^2\ll 1,\label{condsr2}\\
&&\gamma\ll 1. \label{cond}
\ea
Equations \Eq{condsr1} and \Eq{condsr2}, eventually with some variants, are the usual SR limit, while eq. \Eq{cond} is new and highly nontrivial to satisfy, since in $\gamma$ there are infinitely many derivatives of the field. Equations \Eq{condsr1} and \Eq{cond} alone might be sufficient to guarantee the validity of the SR approximation for the SFT tachyon, although one can object that the second one sets an overall constraint on the full SR tower rather than on individual terms. Since the SR approximation alone may result in too weak a condition for the solution of the system, we are forced to the conclusion that in this kind of nonlocal theories an `extended' SR approximation (eqs. \Eq{condsr1}--\Eq{cond}) rather than the usual one is required for consistency.


\subsection{Quasi de Sitter limit}\label{cosds}

Let us briefly explore the consequences of assuming the validity of the approximation given by eq. \Eq{appr} in a quasi de Sitter limit. As regards the final outputs, this is equivalent to state that the infinite series of SR parameters coming from $e^\B f(\phi, \B\phi)$ terms sum up to some finite contribution, eventually decomposed into a leading and a subdominant piece.

In an extended quasi de Sitter regime, the infinite series of derivatives of $H$ are negligible. If the scalar $\tphi$ admits a Fourier decomposition, 
\be
\tphi(t)=\int_{-\infty}^{+\infty}\frac{d\omega}{\sqrt{2\pi}}\,\tphi_\omega e^{i\omega t},
\ee
then the tachyon field can be written as a Gaussian convolution. Defining $\tilde t=t+H\ln\l$, one has (all the integrals below are evaluated on the full real axis)
\ba
\phi(t) &\approx&\int \frac{d\omega}{\sqrt{2\pi}}\,\tphi_\omega\, e^{-c_1\omega^2}e^{i\omega \tilde t}=\int \frac{dt'}{\sqrt{4\pi c_1}}\, e^{-\frac{t'^2}{4c_1}}\int \frac{d\omega}{\sqrt{2\pi}}\,\tphi_\omega e^{i\omega (\tilde t-t')}\nonumber\\
&=& \int \frac{dt'}{\sqrt{4\pi c_1}}\,\tphi(t')\, e^{-(\tilde t-t')^2/(4c_1)}.
\ea
In particular, the \emph{convolution} of $\tphi$ is defined as \cite{MZ}
\be
C[\tphi](t)\equiv \l^{-(2/3)\B}\tphi(t)= \int_{-\infty}^{+\infty} \frac{dt'}{\sqrt{8\pi c_1}}\,\tphi(t')\, e^{-(\tilde t-t')^2/(8c_1)},
\ee
so that eq.~\Eq{teom2} becomes
\be\label{conveom}
(\B-m^2)C[\tphi]=U'.
\ee
The equations of motion in this integral form were studied in \cite{AJ} for the Minkowski case. There the form of the supersymmetric potential was reconstructed by imposing a kink-like ansatz for $\phi(t)$. Note that the iterative method of \cite{vol03} to find kink solutions exploits precisely the integral form of the nonlocal factor (see also \cite{vla05}). 

In general, given a particular potential it is not simple to track even approximate solutions. These can be found by truncating the expansion of the exponential operators. At first order in the small parameter $c_1$,
\be\label{appro}
\tphi^n=(e^{c_1\B}\phi)^n\approx \left(1+nc_1\frac{\B\phi}{\phi}\right)\phi^n.
\ee
This approximation holds if the field does not oscillate too wildly. It was shown in \cite{are04} that this might be the case after some time, at least for kink solutions in Minkowski spacetime. The Friedmann equation and the equation of motion for a tachyon with monomial potential become
\ba
&& H\approx\beta_\t^q(\tilde V-\cO_1)^{q/2},\\
&&3H\dot\phi\left[1-\frac{n-1}{2}(4c_1)\sigma\phi^{n-2}\right]+m^2\phi\left[1-\frac{1}{(-m^2)}\sigma\phi^{n-2}\right]\approx 0.
\ea
A particularly simple example is the bosonic one, eq. \Eq{boso}. Since $4c_1\approx 1.05$, the first equation reduces to $3H\dot\phi\approx\phi$. The authors of \cite{are04,AJ} adopted an approximation similar to eq.~\Eq{appro}, leading to a local action for a ghost field where nonlocal effects are neglected.\footnote{The approximated solutions found in \cite{AJ} are based on a Hamilton--Jacobi equation $\dot H\propto\dot\phi^2$ for a normal field, which agrees with eq.~\Eq{hj1} with $\t=0$ only when one drops the function $\cO_2$. That is, in the opposite case when a phantom behaviour occurs.} This may provide a motivation for the study of cosmological phantom fluids, but in order to retain novel features from the CFT formulation a more refined treatment is required. In fact, the nonlocality of the theory is completely broken, since eq.~\Eq{appro} corresponds to a truncation at finite derivatives. We believe that a study of the cosmological equations of 
motion outside the approximation eq.~\Eq{appro} would be not only interesting but also fundamental for a better understanding of the nonlocal model.


\subsection{Stability and cosmological perturbations}\label{cosst}

Even if we have not find out an exact solution for the cosmological SFT tachyon, we can describe some general features of the associated perturbations. Again, it will be useful to review the Minkowski case. Typically although not always, higher-derivative theories are not ghost-free (well-known counter-examples are $f(R)$ and Gauss--Bonnet gravities) \cite{EW2,sim90,ach94}. Due to the nonlocal nature of the CSFT action, the theory is likely to be unstable already at the classical level.

In Minkowski spacetime, it is easy to verify that the only instability arising from these models is of tachyonic type, but there is no ghost, as it should be since SFT is ghost free by construction. A constant solution of the equation of motion \Eq{teom2} is $\tphi_0^{n-2}=-m^2/\sigma$ for the potentials so far considered. Perturbing the equation of motion around the constant solution, one finds the propagator for $\delta\tphi$ in momentum space ($\B=-k^2$):
\be
\tilde G(-k^2)=[(k^2+m^2)\l^{2k^2/3}-(n-1)m^2]^{-1}.
\ee
For both the bosonic ($m^2=-1$, $n=3$) and supersymmetric ($m^2=-1/2$, $n=4$) tachyon, the propagator has no poles, a feature typical of off-shell fields; see Fig.~\ref{fig1}. This result is possible only for special values of the tachyon mass, and in general (that is, outside string theory) two poles will appear, one of which carrying a ghost. The critical masses above which this happens are $m_{\rm crit}\approx -0.44$ for $n=3$ (at $k^2=-1.47$) and $m_{\rm crit}\approx -0.27$ for $n=4$ (at $k^2=-1.64$). Therefore, in these nonlocal models there is always at least one kind of instability, either tachyonic or ghost-like, with a narrow interval $m_{\rm crit}\leq m<0$ in which both are present. When $m^2=0$, there is a single pole; as noted in \cite{BMS}, the factors $e^\B$ dressing a massless scalar field do not give rise to states different from the tachyonic one, since the resulting propagator does not vanish nor diverges at finite momentum.

\FIGURE{\includegraphics[width=8.6cm]{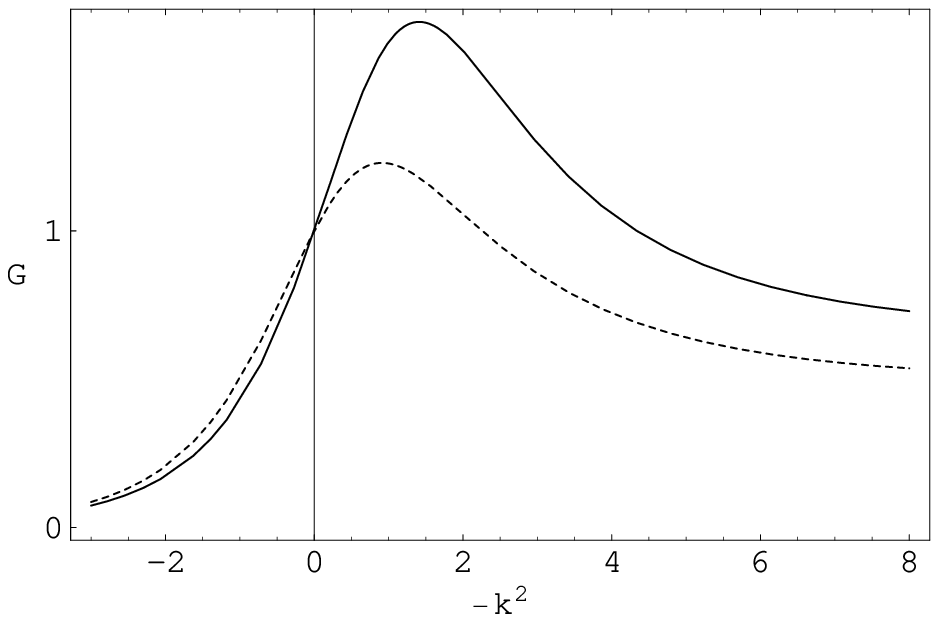}
\caption{\label{fig1}
The propagator $\tilde G(-k^2)$ in momentum space for the bosonic (dashed line) and supersymmetric (solid line) tachyon.}}

In the cosmological case, the tachyon is minimally coupled with gravity and the gravitational sector can be treated separately from the scalar one. $f(R)$ gravities are ghost free when $\p f/\p R>0$ (the proof for the FRW background can be found, e.g., in \cite{HN1,HN2}; the general proof is straightforward via a conformal transformation, but the cited papers are instructive because their calculations lead to the observable power spectra). An expanding accelerating universe with evolution equation \Eq{FRW2} can be described by an effective gravitational action $\cL_{\rm eff}\sim R^{1/q}$. Then, if $\epsilon\ll 1$, $\p f/\p R\sim (2-\t) H^{-\t}>0$ if $\t<2$ (positive $q$), a condition satisfied in general relativity as well as Randall--Sundrum and Gauss--Bonnet braneworlds.

As regards the Klein--Gordon case ($\l=1$), on a FRW background scalar perturbations are governed by the Mukhanov equation
\be\label{psemu}
\p_\tau^2u_{\mathbf k}+\left(|\mathbf{k}|^2+M^2\right)u_{\mathbf k}=0,\\
\ee
where $\tau=\int dt/a$ is conformal time and $u_{\mathbf k}= a\delta\phi_{\mathbf k}$ is the scalar perturbation in the uniform-curvature gauge. The effective mass term $M^2$ is not positive definite in general, and there is the possibility to encounter tachyonic instabilities, even if $m^2>0$. However, the scalar mode is not a ghost as long as the background scalar field is not, too.

If we regard $\phi$ as a test field, that is, which does not contribute to the background, then Eq.~\Eq{psemu} can be derived directly from the perturbed Klein--Gordon equation on the background metric, and the effective mass reads
\be
M_{\rm test}^2=a^2(m^2+U'')-\frac{d_\tau^2a}{a}\approx -(aH)^2(2-3\eta-4\epsilon),
\ee
where we have used $d_\tau^2a/a=(aH)^2(2-\epsilon)$, $m^2+U''\approx 3H^2(\eta+\epsilon)$. In de Sitter ($\epsilon=0=\eta$) it is negative definite, as anticipated. The correct Mukhanov equation, taking into account also metric perturbations, has a mass $M^2\approx M_{\rm test}^2-(aH)^26\epsilon$ to lowest slow-roll order. The extra factor of $6\epsilon$ comes from the scalar modes in the linear perturbations.

The CSFT tachyon has a far richer phenomenology. By perturbing eq.~\Eq{teom2} with $\tphi=\tphi(t)+\delta\tphi(t,{\mathbf x})$, in momentum space and conformal time one has
\ba
&&(\p_\tau^2+|\mathbf{k}|^2)\l^{2k^2/(3a^2)}\tilde u_{\mathbf k}+\tilde M^2_{\mathbf k}\tilde u_{\mathbf k}=0,\\
&& k^2=\p_\tau^2+2\frac{d_\tau a}{a}\p_\tau+|\mathbf{k}|^2,\\
&& \tilde u_{\mathbf k}=a\delta\tphi,\qquad\tilde M^2_{\mathbf k}=a^2(m^2+\tilde U''_{\mathbf k})-\frac{d_\tau^2a}{a},\\
&&\tilde U''_{\mathbf k}\equiv\int \frac{d^4x}{(2\pi)^2}e^{-ik\cdot x}\frac{\delta \tilde U'}{\delta\tphi}\Bigg|_{\tphi(t)}.
\ea
This is the Mukhanov-like equation for a test tachyon rather than the physical one, but it is sufficient for a qualitative comparison. The solution around which one perturbs is no longer constant as in flat spacetime. Therefore the cases $P=\emptyset$ and $P\neq \emptyset$ will give different results;\footnote{Conversely, in \cite{AJK,are04,AJ} the case $P=\emptyset$ was taken to approximate the other one, the field $\tphi$ not oscillating too much. It was checked in \cite{vol03} that this approximation seems good enough for kink-type solutions. So far we cannot say anything about its validity in cosmology when we are far from the almost de Sitter regime.} in the latter situation $M^2_{\mathbf k}$ is in fact a differential operator acting on $\tilde u_{\mathbf k}$. The operators $e^{k^2}$ modify the propagator so that there will be a certain time-dependent range for the effective mass such that ghost modes will appear, as in the Minkowski case. It is clear that the details of this behaviour will strongly depend on the exact solution around which one perturbs the equation of motion.


\subsection{Field redefinition}\label{cosfr}

In the attempt to compare the physics of the SFT tachyon $\phi$ with that predicted in other frameworks, one can try to establish a direct connection between the former and the tachyonic mode of the latters, which we will call $T$. In fact, there is a mapping $f$ between the `well-behaving' tachyon $T$, built via conformal or boundary SFT techniques, and the cubic tachyon. This field redefinition, $\phi(t)=f[T(t)]$, was explicitly constructed in \cite{CST,col04}. 

One can ask if there exists such a mapping even on curved backgrounds. An effective action for $T$, recently under special scrutiny in cosmology, is of DBI type:
\be
\cS_T =-\int d^D xV(T)\sqrt{-g(1+\p_\mu T \p^\mu T)}\,,
\ee
where $V$ is a local potential. The energy density and pressure read, respectively,
\be
\rho(T) = \frac{V(T)}{\sqrt{1-\dot{T}^2}}\,,\qquad p=-V(T)\sqrt{1-\dot{T}^2}\,.
\ee
In Minkowski spacetime, the DBI action does not reproduce exactly the same results as other theories, and it is regarded as an approximated effective model. Also, the DBI action is local (it contains only first-order derivatives), while the CSFT action is not. For these reasons we do not expect to find a relation $\phi \leftrightarrow T$ neither exactly nor of more profound relevance than at the effective level.

Before concluding, however, we give an example of such nonexact relations, which emerges during inflation. In an accelerating regime, the (extended) slow-roll approximation is taken to be valid, and both the DBI and CSFT tachyon behave as normal scalar fields. Therefore, the correspondence between the two is trivial in this limit. To see this, one can write the Hamilton--Jacobi equation \Eq{hjy1}, in units $\beta_\t=1$, as
\ba\label{hjy2}
&& a'H'=-\frac{3q}{2}\,aH^{1+\tilde\t},\\
&& \tilde\t \equiv \t+\frac{\ln (1-\cO_2)}{\ln H}.
\ea
The time variation of the function $\tilde\t$ is
\be
\dot{\tilde\t}=\frac{H}{\ln H}[(\tilde\t-\t)\epsilon-\gamma],
\ee
which vanishes asymptotically in the extended SR limit. Then the SFT tachyon mimics either a normal field, $\tilde\t\approx\t$, or a DBI tachyon, $\tilde\t\approx 2$. 


\section{Discussion}\label{disc}

The cosmological behaviour of the tachyon field in the framework of cubic string field theory is potentially rich of new consequences and worth of investigation. A priori there are several problems affecting the field. First, it turns out that the nonlocal structure of the action is of difficult treatment and sometimes obscure interpretation, in particular when linked to inflation. Second, in the Minkowskian case the rolling solutions, after the above mentioned period of matter behaviour, undergo a  wild oscillatory phase and the interpretation as a matter field breaks down. This might forbid to view the CSFT tachyon as a physical field in cosmology, if neither the Hubble friction nor other mechanisms intervene in smoothing the unstable phase. There is a number of related issues which we have tried to understand and surely would deserve further attention. We list those we believe to be more important and interesting to address.
\begin{itemize}
\item To find approximate cosmological solutions valid even outside the SR approximation. From the lessons gained in the Minkowski case, this task is likely to be carried out with strong help from numerical methods. On the other hand, an exact, complete analytical solution seems, at the present, a goal beyond hope. There are several formalisms which might lead to the right path, as the perturbative one or the $1+1$ Hamiltonian formulation. These should also help to quantize the field in a consistent way.
\item Once one or more homogeneous solutions are found, to discuss the stability of the system around these solutions. This is related to the problem of finding the cosmological perturbation equations. At this stage, is not clear whether the CSFT tachyon would generate isocurvature perturbations or not. In general, the nonadiabatic pressure perturbation $\delta p_\text{nad}\equiv \dot{p}\,[(\delta p/\dot{p})-(\delta\rho/\dot{\rho})]$ vanishes identically for a fluid with a well-defined equation of state $p=p(\rho)$. In the normal or DBI scalar field case, the presence of the inflationary attractor determines univocally the solution (one of the modes decays and can be neglected very early), and the above condition of adiabaticity is satisfied \cite{WMLL}. The existence of such an attractor is guaranteed in an extreme extended SR regime, but as we have shown this case is trivial; in a genuine nonlocal regime it is still to be proved.
\item From the perturbation equations, extract the cosmological observables.
\item Generalize to the case of open and closed universes.
\end{itemize}
All these points concern the above phenomenological model, but there are others of more essential nature, and not less urgent. A solution of the equations of motion will of course depend on the choice for the potential. In Minkowski spacetime, the tachyon solution rapidly converges as the higher string modes are included (and integrated out giving corrections to the tachyon effective potential), so that one has good accuracy already at level 2 \cite{CST}. As noted in \cite{ohm01}, the corrections from the higher-level modes do not typically change the qualitative shape of the effective potential obtained at 0 level, and at most deepen the bottom of the potential. We expect a similar behaviour in the presence of cosmological friction. One might also have the opposite attitude and reconstruct the potential from a given scale factor evolution.

It should be noted that at $(0,0)$ level the tachyonic action is gauge invariant but the corrections to the tachyon potential depend on the gauge choice (while the difference of the potential values between the false and true vacuum is gauge invariant). The results quoted in the introduction are based upon the Siegel gauge, which has been shown to be a reasonable choice for fixing the gauge freedom of the string field action \cite{MS}. When selecting a specific form of the tachyon potential, it would be interesting to look also among other possibilities, some of which might be more suitable for analytic purposes. The ${\cal B}_0$ gauge of \cite{sch05} might be a promising choice in this direction.

Another possibility is to do an analysis without integrating the other particle fields, and include some of them (for instance, the massless gauge boson) as cosmological fields. This step is presumably subdued to the understanding of the tachyon case first.

Let us conclude by quoting Woodard \cite{woo00}: \emph{``In the long struggle of our species to understand the universe it has never once proven useful to invoke a theory which is nonlocal on the most fundamental level.''} Here we have not yet said anything definite about the usefulness of such theories. Nevertheless, they are interesting from a mathematical point of view and, in the context of cosmology, might give rise to radically different phenomena testable through large-scale observations.\footnote{We should note that some nonlocal theories do provide physical pictures that can be constrained by observations. An example is given by noncommutative inflationary models (see \cite{PhD} for related literature, in particular for Brandenberger--Ho models \cite{BH}), which predict a blue-tilted CMB spectrum at large scales. However, there nonlocality acts at the quantum level and in an effective theory, in other words, not at the ``most fundamental level'' of Woodard's statement.\\
By the way, in Brandenberger--Ho models the action measure $\sqrt{-g}$ ($=a^3$ for the FRW metric) is coupled via a * product to the Lagrangian density, which is not the case considered here.} We hope to fulfill the above agenda in the near future.


\acknowledgments{The author thanks I.Ya.~Aref'eva, A. De Felice, V. Forini, G. Grignani, M. Hindmarsh, A.S. Koshelev, A. Liddle, and M. Schnabl for useful discussions. This work is supported by the Angelo della Riccia Foundation.}


\end{document}